\documentclass[aps,pra,superscriptaddress,amsmath,amssymb,preprintnumbers,showpacs,twocolumn]{revtex4}

\usepackage{amssymb}
\usepackage{graphicx}
\usepackage{subfigure}
\usepackage{bm}
\usepackage{color}

\newcommand{\Ket}[1]{\left\vert #1\right\rangle}
\newcommand{\Bra}[1]{\left\langle #1\right\vert}

\newcommand{\BraKet}[2]{\left\langle#1\vert #2\right\rangle}
\newcommand{\KetBra}[2]{\left\vert#1\right\rangle\left\langle#2\right\vert}
\newcommand{\Projector}[1]{\KetBra{#1}{#1}}

\renewcommand{\eqref}[1]{(\ref{#1})} 

\newcommand{\Empty}[1]{ }

\begin{document}

\title{The Role of Temperature in the occurrence of some Zeno Phenomena}

\author{B. Militello}
\affiliation{Dipartimento di Fisica dell'Universit\`{a} di
Palermo, Via Archirafi 36, 90123 Palermo, Italy}
\email{benedetto.militello@unipa.it}

\begin{abstract}
Temperature can be responsible for strengthening effective
couplings between quantum states, determining a hierarchy of
interactions, and making it possible to establish such dynamical
regimes known as Zeno dynamics, wherein a strong coupling can
hinder the effects of a weak one. The relevant physical mechanisms
which connect the structure of a thermal state with the appearance
of special dynamical regimes are analyzed in depth.
\end{abstract}

\maketitle


\section{Introduction}\label{sec:Introduction}

In the last decades a lot of attention has been devoted to the
role of temperature in quantum information and in quantum
mechanics in general. In particular, the detrimental effects of an
environment at finite or even high temperature have been studied
in connection with many aspects of quantum mechanics.
Nevertheless, many studies aimed at singling out robustness of
quantum features even in the presence of high temperature, which
for instance is the case of thermal entanglement. Recently, the
connection between temperature and Quantum Zeno phenomena has also
been explored.

The concept of quantum Zeno effect (QZE)
--- according to its original definition QZE consists in the
inhibition of the dynamics of a physical system due to repeated
\lq\lq pulsed\rq\rq\, measurements~\cite{ref:MishraSudarshan},
which has been demonstrated in various
systems~\cite{ref:Itano1990,ref:Fischer1997}
--- has been progressively generalized. The first generalization
consists in considering continuous measurements meant as
dissipative processes~\cite{ref:Schulman1998}, showing that a
strong decay can be responsible for a dynamical decoupling that
weakens the effects of other interactions~\cite{ref:Panov1999a,
ref:Panov1999b, ref:Militello2011, ref:Scala2010}. Another
important generalization is related to the knowledge that even an
additional coherent coupling can be responsible for diminishing
the effects of other interactions, hence realizing the so called
Zeno Dynamics~\cite{ref:PascazioFacchi2001, ref:Militello2001A,
ref:Militello2001B} and eventually leading to the concept of Zeno
subspaces~\cite{ref:PascazioFacchi2002}. Also pulsed interactions
have been proven to be responsible for the formation of
dynamically invariant subspaces~\cite{ref:PascazioFacchi2004,
ref:PascazioFacchi2008, ref:PascazioFacchi2010}.

Quantum Zeno effect has been exploited to develop applications in
various fields, such as quantum information and
nanotechnologies~\cite{ref:App1, ref:App2, ref:App3}. Moreover,
its importance in connection with fundamental concepts of quantum
mechanics has been widely discussed~\cite{ref:Home1997}.
Nevertheless, Zeno phenomena are not confined to the realm of
quantum mechanics, but, according to Peres, similar occurrences
are possible even in classical mechanics~\cite{ref:Peres1980}.

The connection between Zeno phenomena and temperature has been
investigated by various authors. In 2002 Ruseckas analyzed the
effects of thermalization of a measurement apparatus on the
appearance and intensity of the quantum Zeno effect, showing that
an higher temperature can amplify the power of pulsed measurements
in inducing Zeno phenomena~\cite{ref:Ruseckas2002}. In 2006
Maniscalco {\it et al } have studied the crossover between QZE and
AZE --- Anti-Zeno effect is the acceleration of the dynamics
induced by measurements, usually incoming for
short-but-not-too-short time intervals between subsequent
measurements --- in connection with the temperature of the bath
the system under scrutiny is interacting with, showing that the
QZE-AZE border can be significantly affected by
temperature~\cite{ref:Sabrina2006}. Recently, Scala {\it et al}
have analyzed Landau-Zener (LZ) transitions in the presence of
environmental effects, even in the case of a bath at finite or
high temperature~\cite{ref:ScalaLZ2011}. The analysis, developed
through a master equation approach beyond the secular
approximation, has shown that temperature can be responsible for a
dynamical decoupling, which eventually leads to a Zeno effect. A
subsequent study~\cite{ref:MilitelloTQZE2011} developed in
connection with a different physical situation --- a simplified
version of the previous problem, wherein a time-independent
Hamiltonian has been considered in place of the LZ time-dependent
coupling scheme --- has shown that in some situations temperature
can clearly contribute to induce a Zeno dynamics, weakening the
effects of coherent couplings.

Starting from these results, one could wonder whether in those
cases wherein temperature can influence more than one interactions
it is possible to have temperature-dependent hierarchies of
couplings, driving the appearance of Zeno behaviors. Indeed, for
instance, if two or more interactions are strengthened by
temperature, occurrence of Zeno phenomena could be obstructed,
because of the impossibility of finding a coupling which is much
stronger than others; in other cases temperature can enhance
inhibition of the dynamics. In this paper we analyze the role of
temperature when it affects more than one interactions, discussing
appearance and disappearance of temperature-induced Zeno dynamics,
and studying in depth the underlying physical mechanisms.

The paper is organized as follows. In section
\ref{sec:two_oscillators} we analyze the dynamics of a three-level
system resonantly interacting with two harmonic oscillators
initially prepared in a thermal state. In section
\ref{sec:Detuning_Effects} we extend the scheme by considering
some detuning effects and show an interesting competition between
detuning and temperature in determining Zeno regimes. In section
\ref{sec:Generalization} we give some preliminary results that
pave the way to the analysis of the case where many oscillators
are involved in the coupling scheme. Finally, in section
\ref{sec:conclusions} we give some conclusive remarks.

\section{Zeno dynamics induced by Temperature}\label{sec:two_oscillators}

Let us consider a three-state system interacting with two harmonic
oscillators initially prepared in a thermal state. The interaction
with the oscillators can induce transitions from the upper state
to the intermediate one and from the intermediate to the lowest.
Both interactions are assumed to be resonant (see
Fig.~\ref{fig:two_osc_res}).

From general statements for the appearance of Zeno
dynamics~\cite{ref:PascazioFacchi2002, ref:PascazioFacchi2004,
ref:Militello2001B}, if the strength of the coupling between
$\Ket{2}$ and $\Ket{3}$ is much greater than the strength of the
other coupling, i.e. if $|g_{23}| \gg |g_{12}|$, then transitions
from $\Ket{1}$ to $\Ket{2}$ --- and vice versa --- are hindered.
Therefore, if the atomic system is initially prepared in the state
$\Ket{1}$ then the system does not evolve significantly.

We will show that this behavior occurs when temperature is high
even if the condition $|g_{23}| \gg |g_{12}|$ is not fulfilled,
provided the frequency of the oscillator assisting $2-3$
transitions is much smaller than the frequency of the oscillator
which couples $\Ket{1}$ and $\Ket{2}$.

\subsection{The Hamiltonian Model}

\begin{figure}
\includegraphics[width=0.40\textwidth, angle=0]{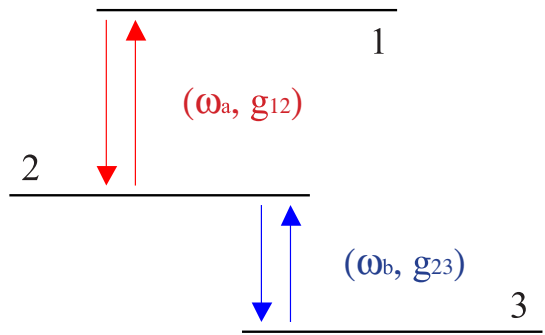}
\caption{(Color online) The upper atomic level (corresponding to
the state $\Ket{1}$) is resonantly coupled to the intermediate
level ($\Ket{2}$), which in turn is resonantly coupled to the
lowest level ($\Ket{3}$).} \label{fig:two_osc_res}
\end{figure}

Consider the following Hamiltonian model ($\hbar=1$):
\begin{eqnarray}
  \nonumber
  && \hat{H} = \sum_k \omega_k \Projector{k} + \omega_a \hat{a}^\dag\hat{a} + \omega_b
  \hat{b}^\dag\hat{b} \\
  \nonumber
  &+& g_{12} (\KetBra{1}{2}\hat{a}+\KetBra{2}{1}\hat{a}^\dag)
  + g_{23} (\KetBra{2}{3}\hat{b}+\KetBra{3}{2}\hat{b}^\dag)\,. \\
  \label{eq:complete_Hamiltonian_model}
\end{eqnarray}
Assuming complete resonance ($\omega_1-\omega_2=\omega_a$ and
$\omega_2-\omega_3=\omega_b$), it clearly corresponds to the
coupling scheme represented in Fig.~\ref{fig:two_osc_res}.

The structure of the Hamiltonian is such that the total number of
excitations,
\begin{equation}
  \hat{N} = 2\Projector{1} + \Projector{2} + \hat{a}^\dag\hat{a} +
  \hat{b}^\dag\hat{b}\,,
\end{equation}
is a constant of the motion. Therefore the Hilbert space is
partitioned into invariant three-dimensional subspaces, each
spanned by a set $\Ket{1, n_a, n_b}$, $\Ket{2, n_a+1, n_b}$ and
$\Ket{3,n_a+1, n_b+1}$, and each one corresponding to a
restriction of the Hamiltonian operator which has the following
form:
\begin{eqnarray}
  \nonumber
  H_{n_a n_b} &=& (n_a\omega_a+n_b\omega_b+\omega_1)\,\mathbb{I}_3\\
  \nonumber
  &\!\!\!\!\!\!+&
  \left(
  \begin{array}{ccc}
    0  & g_{12} \sqrt{n_a+1} & 0 \cr
    \,\, g_{12} \sqrt{n_a+1} \,\, &  0  & \,\, g_{23} \sqrt{n_b+1} \,\, \cr
    0 & g_{23} \sqrt{n_b+1} & 0
  \end{array}
  \right)\,. \\
  \label{eq:restricted_Hamiltonian_rewritten}
\end{eqnarray}
where $\mathbb{I}_3$ is the identity of the $3\times 3$ matrix
space. The \lq restricted\rq\, operator $H_{n_a n_b}$ generates
the \lq restricted\rq\, unitary evolution $U_{n_a n_b} = \exp(-i t
H_{n_a n_b})$.

We now prove that when $g_{23}\sqrt{n_b+1}\gg g_{12}\sqrt{n_a+1}$
the survival probability of the state $\Ket{1, n_a, n_b}$ is
preserved. Preliminarily, consider that one of the three
eigenenergies associated to $H_{n_a n_b}$ is \,
$n_a\omega_a+n_b\omega_b+\omega_1$\,\,
--- according to the analysis in the appendix \ref{App:Matrix3x3},
the matrix in the second line has a vanishing eigenvalue --- and
that the corresponding eigenstate is
\begin{eqnarray}
  \nonumber
  \Ket{n_a\omega_a+n_b\omega_b+\omega_1} &=& \frac{g_{23}\sqrt{n_b+1}}{g}\Ket{1, n_a, n_b} \\
  && \!\!\!\!\!\!\!\!\!\!\!\!\!\!\!\!\!\!\!\!\!\!\!\!\!\!\!\!\!\!
  - \frac{g_{12}\sqrt{n_a+1}}{g}\Ket{3, n_a+1, n_b+1}\,,
\end{eqnarray}
with $g=\left[g_{12}^2(n_a+1)+g_{23}^2(n_b+1)\right]^{1/2}$.

Now, given a positive number $\varepsilon$, provided
\begin{equation}\label{eq:Zeno_Dynamics_Condition}
\frac{g_{23}\sqrt{n_b+1}}{g_{12}\sqrt{n_a+1}} \,\ge\,
\chi_\varepsilon \,\equiv\,
\sqrt{\frac{1+(1-\varepsilon)^{1/4}}{1-(1-\varepsilon)^{1/4}}} \,,
\end{equation}
one easily gets that if the initial state is
$\Ket{\psi(0)}=\Ket{1, n_a, n_b}$ then the population of such
initial state is kept close to unity through all the evolution:
\begin{eqnarray}
  P_{1 n_a n_b}(t)
  &\ge& \sqrt{1-\varepsilon}\,,\qquad \forall t \ge 0\,,
\end{eqnarray}
which can be proven on the basis of the analysis in appendix
\ref{App:Matrix3x3}.

This result can be thought of as the occurrence of a Zeno
dynamics, since the $2-3$ coupling hinders the dynamics induced by
the $1-2$ interaction.

\subsection{Inhibition of Time Evolution}

Let us now consider the system in the following initial state:
\begin{equation}
  \rho(0) = \Projector{1} \otimes \rho_a \otimes \rho_b\,,
\end{equation}
where
\begin{subequations}
\begin{eqnarray}
  \rho_k &=& \mathcal{N}_k\sum_n \exp\left({-\,\omega_k n /k_B
  T}\right) \Projector{n}\\
  \mathcal{N}_k &=& \left[1-\exp\left({-\,\omega_k/k_B T}\right)\right]\,,
\end{eqnarray}
\end{subequations}
with $k=a,b$, are the thermal states of the two oscillators.

The probability to find the three-state system in the initial
state $\Ket{1}$ after a time $t$, irrespectively of the number of
excitations of the harmonic oscillators, is given by:
\begin{eqnarray}\label{eq:Population_General}
  P_1(t)
\nonumber
  &=& \mathcal{N}_a \mathcal{N}_b
  \sum_{n_a} \sum_{n_b} \,
  \exp\left(-\frac{\omega_a n_a + \omega_b n_b}{k_B T} \right)\\
  &\times&
  \left|\Bra{1, n_a, n_b} \hat{U}_{n_a n_b}(t) \Ket{1, n_a, n_b}
  \right|^2\,.
\end{eqnarray}

Condition in Eq.(\ref{eq:Zeno_Dynamics_Condition}) can be
rewritten as
\begin{eqnarray}
  n_b \ge \tilde{n}_b(n_a, \varepsilon)\,,
\end{eqnarray}
with
\begin{eqnarray}
  \nonumber
  \tilde{n}_b(n_a, \varepsilon) &=& \frac{g_{12}^2}{g_{23}^2}\, \chi_\varepsilon^2 \,
  n_a + \frac{\chi_\varepsilon^2 \, g_{12}^2 - g_{23}^2}{g_{23}^2}\\ &\equiv& \alpha_\varepsilon n_a +
  \beta_\varepsilon\,.
\end{eqnarray}

Since all the terms in the right-hand side of
Eq.(\ref{eq:Population_General}) are nonnegative, one has
\begin{eqnarray}
  P_1(t)
\nonumber
  &\ge& \mathcal{N}_a \mathcal{N}_b
  \sum_{n_a} \sum_{n_b^*} \,
  \exp\left(-\frac{\omega_a n_a + \omega_b n_b}{k_B T}\right) \\
  &\times&
  \left|\Bra{1, n_a, n_b} \hat{U}_{n_a n_b}(t) \Ket{1, n_a, n_b}
  \right|^2\,,
\end{eqnarray}
where $n_b^*$ stands for $n_b \ge \tilde{n}_b(n_a,
\,\varepsilon)$. By definition, $|\Bra{1, n_a, n_b} \hat{U}_{n_a
n_b}(t) \Ket{1, n_a, n_b}|^2 \ge \sqrt{1-\varepsilon}$\, for any
$n_b^*$, and then one has:
\begin{widetext}
\begin{eqnarray}
  P_1(t)
\nonumber
  &\ge& \mathcal{N}_a \mathcal{N}_b
  \sum_{n_a} \sum_{n_b^*} \,
  \exp\left(-\frac{\omega_a n_a + \omega_b n_b}{k_B T}\right)
  \sqrt{1-\varepsilon}
%
\nonumber
  \,\, = \,\, \mathcal{N}_a \sum_{n_a}
  \, \exp\left[-\frac{\omega_a n_a + \omega_b [[\tilde{n}_b(n_a,\,\varepsilon)]]}{k_B T}\right] \, \sqrt{1-\varepsilon} \\
\nonumber
  &\ge& \mathcal{N}_a \sum_{n_a}
  \, \exp\left[-\frac{\omega_a n_a + \omega_b [\tilde{n}_b(n_a,\,\varepsilon)+1]}{k_B T}\right] \, \sqrt{1-\varepsilon} \\
  &=& \exp\left[-(\beta_\varepsilon+1)\omega_b/(k_B T)\right]\frac { 1 - \exp\left[-\,\omega_a/(k_B T)\right] }{ 1 - \exp\left[-\left(\omega_a+\alpha_\varepsilon \omega_b\right) / (k_B T)\right] }
  \, \sqrt{1-\varepsilon} \,,
  \label{eq:P1_Lower_Bound_I}
\end{eqnarray}
\end{widetext}
where the symbol $[[x]]$ indicates the first integer number larger
than or equal to $x$, and the relations $x+1\ge [[x]]$ and
$\sum_{n=s}^{\infty} e^{-nx}=e^{-sx}/(1-e^{-x})$ have been used.

For any given value of $\varepsilon$, positive and smaller than
unity, in the limit of very high temperature, one can assume
$\exp[-(\beta_\varepsilon+1)\omega_b/(k_B T)] \approx 1$ and
$1-\exp[x/(k_B T)] \approx x/(k_B T)$, so that one gets
\begin{eqnarray}
  \label{eq:P1_Lower_Bound_II}
  P_1(t) &\gtrsim &
  \frac{\sqrt{1-\varepsilon}}{1+\eta\,\chi^2_\varepsilon}\,,
\end{eqnarray}
with
\begin{eqnarray}\label{eq:Definition_Eta}
\eta =
\frac{g_{12}^2}{g_{23}^2}\frac{\omega_b}{\omega_a}\,.
\end{eqnarray}

Therefore, in the limit of high temperature the survival
probability becomes higher and higher for any $t$, approaching
unity in the limit $\eta\rightarrow 0$. In fact, for $\eta \ll 1$,
by choosing $\varepsilon = \sqrt{\eta}$, since $\chi^2_\varepsilon
\approx 8/\varepsilon$ for very small $\varepsilon$, one gets
$P_1(t) \gtrsim 1-(17/2)\sqrt{\eta} \rightarrow 1$. The limit
$\eta \rightarrow 0$ can be reached in different ways: two
possibilities are assuming $|g_{12}| \ll |g_{23}|$ or assuming
$\omega_b/\omega_a \ll 1$. On this basis, one can assert that in
each of the two limits, $|g_{12}/g_{23}| \rightarrow 0$ or
$\omega_b/\omega_a \rightarrow 0$, it turns out that $P_1(t)
\rightarrow 1$ for any $t$, provided the temperature is high
enough.

It is worth emphasizing that in order to achieve the limit in
(\ref{eq:P1_Lower_Bound_II}) one must legitimate truncation of the
series expansions of the exponentials involved, which is possible
under the assumption that the following conditions are fulfilled:
\begin{subequations}
\begin{eqnarray}
(\beta_\varepsilon+1)\omega_b / (k_B T) \ll 1\,, \\
(\omega_a+\alpha_\varepsilon \omega_b) / (k_B T) \ll 1\,.
\end{eqnarray}
\end{subequations}
These two inequalities define the limit of high temperature for any fixed value of $\varepsilon$.

\begin{figure}
\subfigure[]{\includegraphics[width=0.40\textwidth,
angle=0]{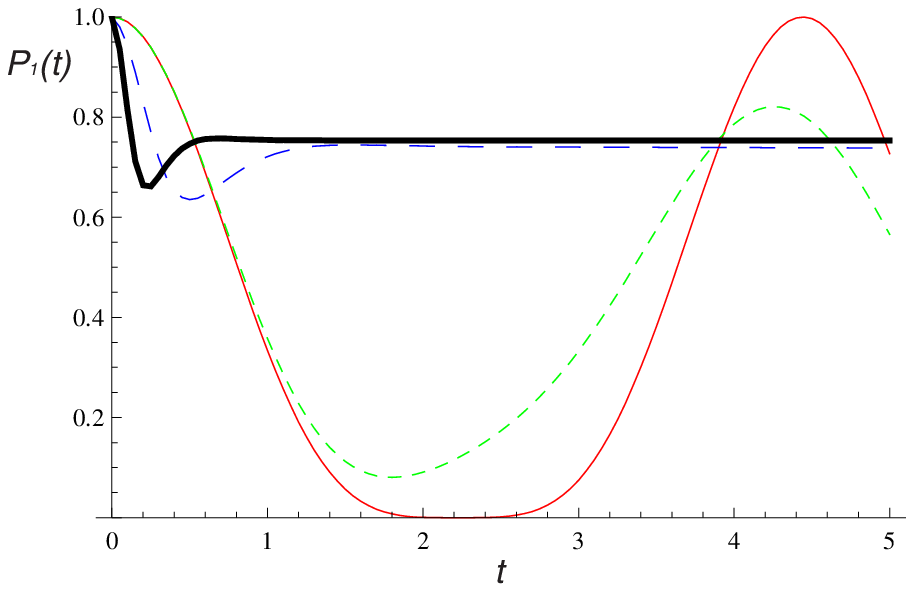}}
\subfigure[]{\includegraphics[width=0.40\textwidth,
angle=0]{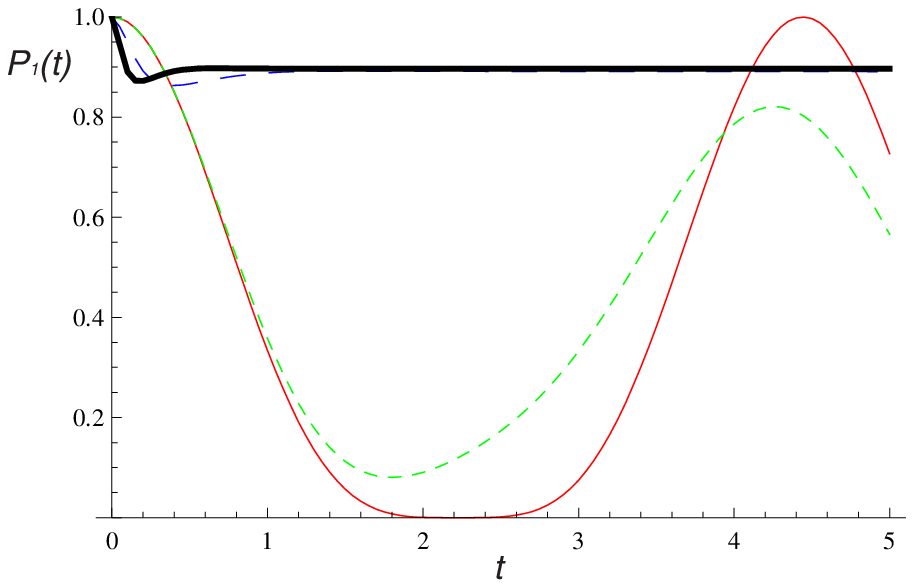}} \caption{(Color online) The
survival probability of the atomic state $\Ket{1}$ as a function
of time, for different values of the relevant parameters. (a)
$\omega_a/\omega_b=10$, $g_{12}/\omega_b=g_{23}/\omega_b = 1$, for
different values of the temperature: $T/\omega_b=0.1$ (red solid
line), $T/\omega_b=1$ (green dotted line), $T/\omega_b=50$ (blue
dashed line), $T/\omega_b=250$ (black bold line). (b)
$\omega_a/\omega_b=50$, $g_{12}/\omega_b=g_{23}/\omega_b = 1$, for
different values of the temperature: $T/\omega_b=0.1$ (red solid
line), $T/\omega_b=1$ (green dotted line), $T/\omega_b=250$ (blue
dashed line), $T/\omega_b=1250$ (black bold line). }
\label{fig:evolutions_two_osc_res_system}
\end{figure}

In Fig.~\ref{fig:evolutions_two_osc_res_system} are shown the
evolutions of the survival probability for different temperatures
and for different ratios of the frequencies of the two
oscillators. It is evident that the higher the temperature the
higher the survival probability at any time. Moreover, the smaller
the ratio $\omega_b/\omega_a$ the higher the asymptotic value of
$P_1(t)$.

It is worth noting the difference with the case analyzed in the
previous paper on Zeno subspaces induced by
temperature~\cite{ref:MilitelloTQZE2011}. In fact, in that case
the limit of the survival probability for $T\rightarrow\infty$ is
unity, while in the present case the survival probability does not
tend toward unity, unless assuming $\omega_b/\omega_a\rightarrow
0$ (or $|g_{12}| / |g_{23}| \rightarrow 0$). This difference can
be explained analyzing the physical mechanism on the basis of
these Zeno-like behaviors induced by temperature.  Indeed, the key
points are two: first, in each invariant subspace the strength of
the coupling with an oscillator grows up as the number of
excitations increases; second, the higher the temperature the more
the subspaces with high excitation numbers are populated. Now, in
the case of Ref.~\cite{ref:MilitelloTQZE2011} the interaction
$1-2$ is not assisted by an harmonic oscillator and therefore it
is not strengthened by temperature as the interaction $2-3$ is. As
a consequence, an higher temperature makes the ratio between the
two coupling strengths tend toward zero in most of the subspaces.
On the contrary, in the present case, both couplings, $1-2$ and
$2-3$, are effectively enhanced by temperature, and one needs
$\omega_b \ll \omega_a$ in order to make the coupling $2-3$
prevail in the majority of the subspaces. Indeed, $\omega_b \ll
\omega_a$ means that excited states of the oscillator $b$ are more
populated than the corresponding states of oscillator $a$, at the
same temperature. If this condition (or the alternative one, i.e.,
$|g_{12}| / |g_{23}|\ll 1$) is not properly fulfilled, occurrence
of Zeno dynamics is attenuated.

\subsection{Low vs High Temperature}

In order to better understand the role of temperature, it could be
useful to compare the behaviors of the survival probability at low
and high temperature.

In fact, in the zero-temperature case, the only block involved in
the dynamics is the one related to $n_a=n_b=0$, and the inhibition
of the dynamics of $\Ket{1,0,0}$ is determined by the condition
$|g_{23}|\gg |g_{12}|$, so that it is essentially due to a
significant difference between the strengths of the two couplings.
On the contrary, when the high-temperature limit is considered,
the parameter determining the minimum value of the survival
probability is $\eta$, as defined in
Eq.~(\ref{eq:Definition_Eta}), which involves both the coupling
strengths and the frequencies of the oscillators. As already
discussed before, a smaller frequency implies an higher degree of
population of the excited levels of the oscillator, and eventually
a stronger coupling in most of the Hilbert space.

\begin{figure}
\includegraphics[width=0.40\textwidth, angle=0]{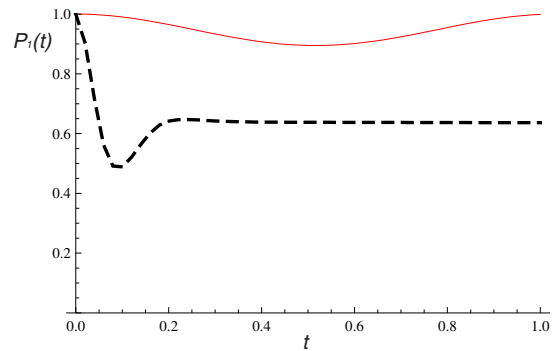}
\caption{(Color online) Survival probability of the atomic state
$\Ket{1}$ at low and high temperature. The relevant quantities are
$\omega_b/\omega_a=10$, $g_{12}/\omega_a=1$, $g_{23}/g_{12} = 6$;
temperature: $T/\omega_a=0.1$ (red solid line), $T/\omega_a=250$
(black dashed line).} \label{fig:low_vs_high_Temp}
\end{figure}

On this basis, one finds that Zeno dynamics is possible even in
the case $|g_{23}|<|g_{12}|$, provided $\omega_b \ll \omega_a$ and
temperature high enough. On the contrary, temperature can produce
the opposite effect, determining disappearance of Zeno dynamics.
As an example, consider the case $|g_{23}|\gg |g_{12}|$, which
implies Zeno dynamics at zero temperature, and $\omega_b >
\omega_a |g_{23}/g_{12}|^2 \gg \omega_a$, which renders $\eta$
large enough to invalidate inhibition of the dynamics. In
Fig.~\ref{fig:low_vs_high_Temp} it is shown a case wherein by
increasing temperature one lowers the survival probability at any
time.

\section{Detuning effects}\label{sec:Detuning_Effects}

\begin{figure}
\includegraphics[width=0.40\textwidth, angle=0]{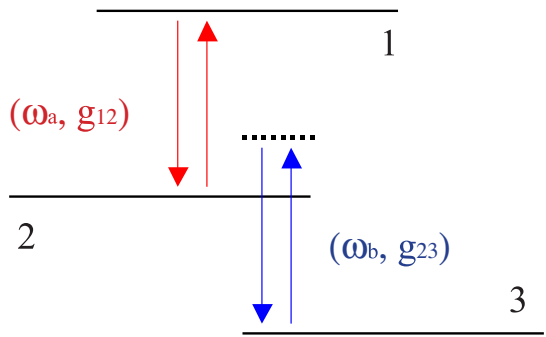}
\caption{(Color online) The upper atomic level ($\Ket{1}$) is
resonantly coupled to the intermediate level ($\Ket{2}$), which in
turn is non resonantly coupled to the lowest level ($\Ket{3}$).}
\label{fig:second_system}
\end{figure}

Let us now consider again the Hamiltonian in
Eq.(\ref{eq:complete_Hamiltonian_model}) but without the resonance
conditions (see Fig.~\ref{fig:second_system}). Generally speaking
the presence of a detuning causes a diminishing of the interaction
strength, so that the presence of a detuning in the $2-3$
transition will make us analyze the competition between detuning
and temperature in connection with the occurrence of Zeno
dynamics. On the contrary, the presence of a detuning in the $1-2$
transition would simply hinder the evolution of the initial atomic
state ($\Ket{1}$) even in the absence of the second oscillator and
relevant coupling. Therefore this analysis would be less
interesting. Hence, for the sake of simplicity, let us assume that
the transition $1-2$ is resonant with the relevant oscillator
($\omega_a=\omega_1-\omega_2$), while the transition $2-3$ is
assumed not to be resonant with the second oscillator:
$\Delta=\omega_b-(\omega_2-\omega_3)\not=0$. The restriction of
the Hamiltonian in the generic invariant subspace corresponding to
$n_a$ and $n_b$ is given by the following matrix:
\begin{eqnarray}
  \nonumber
  H_{n_a n_b} &=& (n_a\omega_a+n_b\omega_b+\omega_1)\mathbb{I}_3\\
  \nonumber
  &\!\!\!\!\!\!+&
  \left(
  \begin{array}{ccc}
    0  & g_{12} \sqrt{n_a+1} & 0 \cr
    \,\, g_{12} \sqrt{n_a+1} \,\, &  0  & \,\, g_{23} \sqrt{n_b+1} \,\, \cr
    0 & g_{23} \sqrt{n_b+1} & \Delta
  \end{array}
  \right)\,. \\
  \label{eq:restricted_Hamiltonian_detuning}
\end{eqnarray}

According with the analysis in appendix \ref{App:Matrix3x3}, in
order to have inhibition of the dynamics of the state
$\Ket{1,n_a,n_b}$ --- making it a quasi-eigenstate of the
Hamiltonian --- one needs that $g_{23}\sqrt{n_b+1}$ is much grater
than $g_{12}\sqrt{n_a+1}$ and greater than $\Delta$. A large value
of detuning can obstacle the appearance of Zeno dynamics, but
larger values of temperature can establish again such a regime by
populating more and more those subspaces wherein
$g_{23}\sqrt{n_b+1}$ is larger (and even much larger) than
$\Delta$.

\begin{figure}
\subfigure[]{\includegraphics[width=0.40\textwidth,
angle=0]{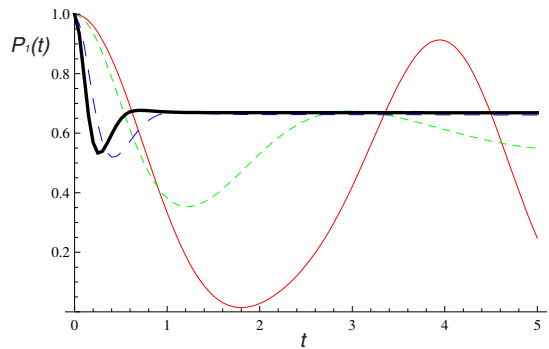}}
\subfigure[]{\includegraphics[width=0.40\textwidth,
angle=0]{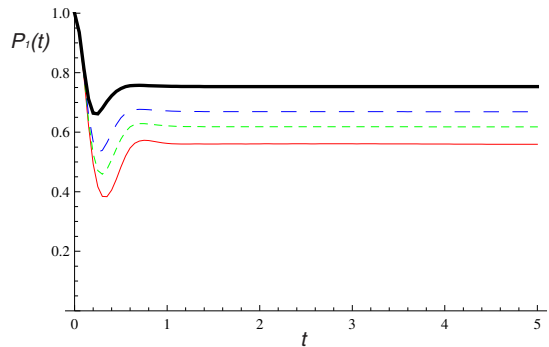}} \caption{(Color online) The
survival probability of the atomic state $\Ket{1}$ as a function
of time, for different values of the relevant parameters. (a)
$\omega_a/\omega_b=10$, $g_{12}/\omega_b=g_{23}/\omega_b = 1$,
$\Delta/\omega_b=1$, for different values of the temperature:
$T/\omega_b=0.1$ (red solid line), $T/\omega_b=10$ (green dotted
line), $T/\omega_b=100$ (blue dashed line), $T/\omega_b=250$
(black bold line). (b) $\omega_a/\omega_b=10$,
$g_{12}/\omega_b=g_{23}/\omega_b = 1$, $k_B T/\omega_b = 250$, for
different values of the detuning: $\Delta/\omega_b=4$ (red solid
line), $\Delta/\omega_b=2$ (green dotted line),
$\Delta/\omega_b=1$ (blue dashed line), $\Delta=0$ (black bold
line).} \label{fig:evolutions_detuning}
\end{figure}

In Fig.~\ref{fig:evolutions_detuning} it is shown the behavior of
the population of the atomic state $\Ket{1}$ as a function of time
but for different values of temperature and detuning. From
Fig.~\ref{fig:evolutions_detuning}a we learn that even in the
presence of non vanishing detuning there is an increase of the
survival probability when temperature becomes higher.
Nevertheless, comparing the bold line in this figure with the one
in Fig.~\ref{fig:evolutions_two_osc_res_system}a, we see that the
asymptotic value of the population is slightly smaller in the
presence of detuning, in spite of the fact that the temperature is
the same.

In Fig.~\ref{fig:evolutions_detuning}b it is shown that, for a
fixed value of temperature, a diminishing of the detuning produces
an increase of the asymptotic population, as expected.

\section{Attempt of Generalization}\label{sec:Generalization}

In view of a possible study of interactions between a few-level
system and its environment, the previous analysis could deserve a
generalization to the case wherein many harmonic oscillators
couple both transitions, $1-2$ and $2-3$. Since we are interested
to the case of quite different Bohr frequencies, we can consider
two independent sets of oscillators. The relevant Hamiltonian is
given by,
\begin{eqnarray}
  \nonumber
  \hat{H} &=& \sum_k \omega_n \Projector{n} + \sum_{k=1}^p \nu_k \hat{a}_k^\dag\hat{a}_k +
  \sum_{l=1}^q\mu_l \hat{b}_l^\dag\hat{b}_l \\
  \nonumber
  &+& \sum_{k=1}^p g^{(12)}_k
  (\KetBra{1}{2}\hat{a}_k+\KetBra{2}{1}\hat{a}_k^\dag) \\
  \nonumber
  &+& \sum_{l=1}^q g^{(23)}_l (\KetBra{2}{3}\hat{b}_l+\KetBra{3}{2}\hat{b}_l^\dag)\,, \\
  \label{eq:NxM_complete_Hamiltonian_model}
\end{eqnarray}
which conserves the following quantity:
\begin{eqnarray}
  \hat{N}=2\Projector{1}+\Projector{2}+\sum_{k=1}^p\hat{a}_k^\dag\hat{a}_k +
  \sum_{l=1}^q\hat{b}_l^\dag\hat{b}_l\,.
\end{eqnarray}

Therefore the Hilbert space is partitioned into invariant
subspaces ${\cal H}_{\vec{n}_p\vec{m}_q}$ of dimension $\dim{\cal
H}_{\vec{n}_p\vec{m}_q}=1+p+pq$ and spanned by the following
vectors: $\Ket{1,\vec{n}_p,
\vec{m}_q}\equiv\Ket{1,n_1,...n_p,m_1,...m_q}$; \,\,
$(n_k+1)^{-1/2}\hat{a}_k^\dag\Ket{2,\vec{n}_p, \vec{m}_q}$ with
$k=1,...p$; \,\,
$(n_k+1)^{-1/2}(m_l+1)^{-1/2}\hat{a}_k^\dag\hat{b}_l^\dag\Ket{3,\vec{n}_p,
\vec{m}_q}$, with $k=1,...p$ and $l=1,...q$.

The restriction of the Hamiltonian to each subspace has the
following structure:
\begin{widetext}
\begin{equation}
  B_{p\times q} = \left(
  \begin{array}{ccccccccccccccc}
    \delta_0 & \alpha_1   & \alpha_2   & ...     & \alpha_p   & 0          & 0          & 0          & 0          & ...    & 0          & ...    & 0          & ...    & 0          \cr
    \alpha_1 & \delta_1   & 0          & ...     & 0          & \beta_{11} & ...        & \beta_{1q} & 0          & ...    & 0          & ...    & 0          & ...    & 0          \cr
    \alpha_2 & 0          & \delta_2   & ...     & 0          & 0          & ...        & 0          & \beta_{21} & ...    & \beta_{2q} & ...    & 0          & ...    & 0          \cr
    \vdots   & \vdots     & \vdots     & \ddots  &            & \vdots     &            &            &            &        &            &        &            &        &            \cr
    \alpha_p & 0          & 0          &         & \delta_p   & 0          & ...        & 0          & 0          & ...    & 0          & ...    & \beta_{p1} & ...    & \beta_{pq} \cr
    0        & \beta_{11} & 0          & ...     & 0          & \delta_{11}&            & 0          & 0          & ...    & 0          & ...    & 0          & ...    & 0          \cr
    \vdots   & \vdots     & \vdots     &         & \vdots     & ...        & \ddots     & \vdots     & 0          & ...    & 0          & ...    & 0          & ...    & 0          \cr
    0        & \beta_{1q} & 0          & ...     & 0          & 0          & ...        & \delta_{1q}& 0          & ...    & 0          & ...    & 0          & ...    & 0          \cr
    0        & 0          & \beta_{21} & ...     & 0          & 0          & ...        & 0          & \delta_{21}& ...    & 0          & ...    & 0          & ...    & 0          \cr
    \vdots   & \vdots     & \vdots     &         & \vdots     & \vdots     &            & 0          & 0          & \ddots & 0          & ...    & 0          & ...    & 0          \cr
    0        & 0          & \beta_{2q} & ...     & 0          & 0          & ...        & 0          & 0          & ...    & \delta_{2q}& ...    & 0          & ...    & 0          \cr
    \vdots   & \vdots     & \vdots     &         & \vdots     & \vdots     &            & \vdots     & \vdots     & ...    & \vdots     & \ddots & \vdots     & ...    & \vdots     \cr
    0        & 0          & 0          & ...     & \beta_{p1} & 0          & ...        & 0          & 0          & ...    & 0          & ...    & \delta_{p1}& ...    & 0          \cr
    \vdots   & \vdots     & \vdots     &         & \vdots     & \vdots     &            & 0          & 0          &        & 0          & ...    & 0          & \ddots & 0          \cr
    0        & 0          & 0          & ...     & \beta_{pq} & 0          & ...        & 0          & 0          & ...    & 0          & ...    & 0          & ...    & \delta_{pq}\cr
  \end{array}
  \right)\,,
\end{equation}
\end{widetext}
where the only non vanishing terms in a line or column are between
two symbols different from \lq\,$0$\rq\, in that line or column.
In spite of this fact, the matrix is rather involved and difficult
to study.

In the very special case of $p=1$, such structure becomes quite
simple:
\begin{equation}\label{eq:1xq_Matrix}
  B_{1\times q} = \left(
  \begin{array}{cccccc}
    \delta     & \alpha    & 0           & 0           & ...    & 0           \cr
    \alpha     & \delta_1  & \beta_1     & \beta_2     & ...    & \beta_M     \cr
    0          & \beta_1   & \delta_{11} & 0           & ...    & 0           \cr
    0          & \beta_2   & 0           & \delta_{12} & ...    & 0           \cr
    \vdots     & \vdots    & \vdots      & \vdots      & \ddots & \vdots      \cr
    0          & \beta_M   & 0           & 0           & ...    & \delta_{1M} \cr
  \end{array}
  \right)\,,
\end{equation}
where $\alpha=\alpha_1$ and $\beta_l=\beta_{1l}$. In the generic
subspace ${\cal H}_{\vec{n}_p\vec{m}_q}$ one has
$\alpha=g_1^{(12)}(n_1+1)^{1/2}$ and
$\beta_l=g_l^{(23)}(m_l+1)^{1/2}$.

The matrix in Eq.~(\ref{eq:1xq_Matrix}) has the same structure of
the matrix analyzed in the Appendix B of
Ref.~\cite{ref:MilitelloTQZE2011}. On the basis of that analysis,
we claim that when the system starts form $\Ket{1,n_1,\vec{m}_q}$
and $(\sum_l \beta_l^2)^{1/2} \gg |\alpha|$ then the dynamics is
hindered. Therefore, focusing our attention on the population of
the atomic state $\Ket{1}$ irrespectively of the number of
excitations of the oscillators, by reasoning as in the previous
sections, we can say that the evolution of such population turns
out to be hindered when the frequencies $\mu_l$ are smaller
--- preferably much smaller --- than $\nu\equiv\nu_1$. Such a
condition on the frequencies makes the previous assertion true
even if, for example, $|g_1^{(12)}|^2 \sim \sum_l|g_l^{(23)}|^2$.

\section{Discussion}\label{sec:conclusions}

The analysis developed in this paper shows in a clear way the role
of the temperature in determining the occurrence of Zeno dynamics.

In particular, we have considered a few-level system prepared in
an excited state and interacting with a set of harmonic
oscillators initially prepared in thermal states at the same
temperature. We have pointed out that an higher temperature
implies a more significant role of the subspaces of the Hilbert
space characterized by higher numbers of excitations, which in
turn imply stronger effective couplings. The smaller the frequency
of an oscillator the more its excited states are involved in the
dynamics, for a given temperature. Therefore, the presence of a
coupling between the lowest state $\Ket{3}$ and the intermediate
one $\Ket{2}$ can significantly weaken the effects of the $1-2$
coupling, especially when the Bohr frequency of the $2-3$
transition is sufficiently small. This is a significant difference
between high-temperature and low-temperature limit, since in the
latter case the role of the frequencies is absent and only
coupling strengths are important.

This analysis is the prosecution of the study developed in
Ref.~\cite{ref:MilitelloTQZE2011} where the $1-2$ coupling was
induced by an external field assumed not to be affected by
temperature. In the present work, instead, we have considered the
competition between the growing of both the couplings, $1-2$ and
$2-3$, and the role of temperature and Bohr frequencies in making
one of them prevail.

A further analysis aimed at considering a set of harmonic
oscillators driving each atomic transition --- similar to the one
performed in the previous paper --- would be appropriate. Some
preliminary results have been reported in section
\ref{sec:Generalization}.

Another point that could deserve attention is the role of possible
counter rotating terms in the interaction between the few-level
system and the bosonic part of the total system.

\section{Acknowledgements}

The Author wishes to thank Antonino Messina and Matteo Scala for
carefully reading the manuscript.

\appendix

\section{} \label{App:Matrix3x3}

In this appendix we prove that a matrix of the following form
(written in the basis $\Ket{A}$, $\Ket{B}$, $\Ket{C}$):
\begin{eqnarray}
  M=
  \left(
  \begin{array}{ccc}
     0  & \alpha_{1} & 0 \cr
    \alpha_{1}^* &  0  & \,\, \alpha_{2} \,\, \cr
     0  & \alpha_{2}^*  & \Delta
  \end{array}
  \right)\,,
\end{eqnarray}
has an eigenvalue whose corresponding eigenstate becomes closer to
$\Ket{A}$ as the modulus of the coupling constant $\alpha_2$ grows
up.

Let us first consider the special situation $\Delta=0$, in which
case we can directly check that $\lambda=0$ is an eigenvalue and
that the corresponding eigenstate is
\begin{equation}
  \Ket{0} = \frac{\alpha_2}{\alpha}\Ket{A} - \frac{\alpha_1^*}{\alpha}\Ket{C}\,,\qquad
  \alpha=\sqrt{|\alpha_1|^2+|\alpha_2|^2}\,\,,
\end{equation}
from which one deduces that $|\BraKet{A}{0}|=|\alpha_2/\alpha|$
tends toward unity in the limit $|\alpha_1|\ll|\alpha_2|$.

Concerning the general case $\Delta\not=0$, the secular polynomial
$P(\lambda)=\det(M-\lambda\mathbb{I}_3)$ is:
\begin{eqnarray}
  P(\lambda) = -\lambda^3 + \Delta\lambda^2 +
  (|\alpha_1|^2+|\alpha_2|^2)\lambda - |\alpha_1|^2\Delta\,.
\end{eqnarray}

It turns out that
\begin{subequations}
\begin{eqnarray}
  && P(0) = - |\alpha_1|^2\Delta\,,
\end{eqnarray}
and
\begin{eqnarray}
  \nonumber
  && P\left(\frac{|\alpha_1^2|}{|\alpha_2^2|}\Delta\right) = \frac{|\alpha_1|^4}{|\alpha_2|^2}\Delta +
  \left(\left|\frac{\alpha_1}{\alpha_2}\right|^4-\left|\frac{\alpha_1}{\alpha_2}\right|^6\right)\Delta^3\,,\\
\end{eqnarray}
\end{subequations}
which for $\Delta\not=0$ and $|\alpha_1|<|\alpha_2|$ implies a
change of sign. Therefore there is an eigenvalue $\eta$ of modulus
$|\eta| < |\alpha_1^2\Delta/\alpha_2^2|$. The relevant eigenvector
is:
\begin{subequations}
\begin{eqnarray}
  \Ket{\eta} = \mathcal{P}_\eta
  \left(\Ket{A}+\frac{\eta}{\alpha_1}\Ket{B}-\frac{\eta\,\alpha_2^*}{(\Delta-\eta)\,\alpha_1}\Ket{C}\right)\,,
\end{eqnarray}
with
\begin{eqnarray}
  \mathcal{P}_\eta = \left[1 + \frac{|\eta|^2}{|\alpha_1|^2} + \frac{|\eta|^2 |\alpha_2|^2}{|\Delta-\eta|^2\,|\alpha_1|^2}\right]^{-1/2}
  \,.
\end{eqnarray}
\end{subequations}

Now, assume $|\Delta| < |\alpha_2|$ and $|\alpha_1|\ll|\alpha_2|$
(which in turn implies $|\eta| \ll |\Delta|$ and then
$|\Delta/(\Delta-\eta)|<2$), one finds that $|\eta/\alpha_1| <
|\alpha_1/\alpha_2|\times|\Delta/\alpha_2|<|\alpha_1/\alpha_2|$
and $\left|\eta^2 \alpha_2^2/[(\Delta-\eta)^2\,\alpha_1^2]\right|
< |\alpha_1/\alpha_2|^2\times|\Delta/(\Delta-\eta)|^2 <
2|\alpha_1/\alpha_2|^2$. Then, under the previous hypotheses one
has $|\BraKet{A}{\eta}|=\mathcal{P}_\eta \rightarrow 1$.

We conclude reminding that (see appendix A of
Ref.\cite{ref:MilitelloTQZE2011}) if the system is prepared in the
initial state $\Ket{\psi(0)}=\Ket{A}$, then the survival
probability has a non vanishing lower bound:
$P_A(t)=|\BraKet{A}{\psi(t)}|^2 \ge (2|\BraKet{A}{\eta}|^2-1)^2$,
which of course is close to unity when $|\BraKet{A}{\eta}| \approx
1$.


\end{document}